\documentclass[11pt]{article}
\usepackage{ifpdf,latexsym,graphicx,url}
\usepackage{times}

\title{Bidimensionality, Map Graphs, and Grid Minors}
\author{Erik D. Demaine%
  \thanks{MIT Computer Science and Artificial Intelligence Laboratory,
        32 Vassar Street, Cambridge, MA 02139, U.S.A.,
        \{edemaine, hajiagha\}@mit.edu}
\and MohammadTaghi Hajiaghayi\footnotemark[1]}
\date{}

\newif\ifabstract
\abstracttrue
\newif\iffull
\ifabstract \fullfalse \else \fulltrue \fi

\usepackage
  [breaklinks,bookmarks,bookmarksnumbered,bookmarksopen,bookmarksopenlevel=2]
  {hyperref}
{\makeatletter \hypersetup{pdftitle={\@title}}}
\advance \topmargin by -\headheight
\advance \topmargin by -\headsep
\evensidemargin \oddsidemargin
\ifabstract
\fi

\let\latexcite=\cite
\def\cite{\nolinebreak\latexcite}
\let\latexref=\ref
\def\ref{\nolinebreak\latexref}

{\makeatletter
 \gdef\xxxmark{%
   \expandafter\ifx\csname @mpargs\endcsname\relax % in minipage?
     \expandafter\ifx\csname @captype\endcsname\relax % in figure/caption?
       \marginpar{xxx}% not in a caption or minipage, can use marginpar
     \else
       xxx % notice trailing space
     \fi
   \else
     xxx % notice trailing space
   \fi}
 \gdef\xxx{\@ifnextchar[\xxx@lab\xxx@nolab}
 \long\gdef\xxx@lab[#1]#2{{\bf [\xxxmark #2 ---{\sc #1}]}}
 \long\gdef\xxx@nolab#1{{\bf [\xxxmark #1]}}
 \long\gdef\xxx@lab[#1]#2{}\long\gdef\xxx@nolab#1{}%
}

{\makeatletter \gdef\fps@figure{!htbp}}

\newtheorem{theorem}{Theorem}
\newtheorem{lemma}[theorem]{Lemma}
\newtheorem{proposition}[theorem]{Proposition}
\newtheorem{corollary}[theorem]{Corollary}

\newtheorem{conjecture}[theorem]{Conjecture}

\newenvironment{proof}{\noindent\textbf{Proof: }\ignorespaces}
  {\hspace*{\fill}$\Box$\medskip}

\def\twodots{.\,.\,}

\def\bag{\mathcal B}
\def\tw{\mathop{\rm tw}\nolimits}
\def\maxdeg{\Delta}

\begin{document}
\maketitle

\begin{abstract}
  In this paper we extend the theory of bidimensionality to two families of
  graphs that do not exclude fixed minors: map graphs and power graphs.
  In both cases we prove a polynomial relation between the treewidth of a
  graph in the family and the size of the largest grid minor.
  These bounds improve the running times of a broad class of
  fixed-parameter algorithms.
  Our novel technique of using approximate max-min relations between
  treewidth and size of grid minors is powerful, and we show how it
  can also be used, e.g., to prove a linear relation between the treewidth
  of a bounded-genus graph and the treewidth of its dual.
\end{abstract}

\section{Introduction}

The newly developing theory of bidimensionality,
developed in a series of papers
\cite{DHT02,jcss,mapgraphs,diamtw,hminorfree,apexfree,localtw,boundedgenus,gridminors,genapprox},
provides general techniques
for designing efficient fixed-parameter algorithms and
approximation algorithms for NP-hard graph problems in broad classes of
graphs.  This theory applies to graph problems that are \emph{bidimensional}
in the sense that (1)~the solution value for $r \times r$
``grid-like'' graphs grows with~$r$, typically as $\Omega(r^2)$, and
(2)~the solution value goes down when contracting edges and optionally
when deleting edges (i.e., taking minors).  Examples of such problems include
feedback vertex set, vertex cover, minimum maximal matching, face cover,
a series of vertex-removal parameters, dominating set, edge dominating set,
$R$-dominating set, connected dominating set, connected edge dominating set,
connected $R$-dominating set, and unweighted TSP tour
(a walk in the graph visiting all vertices).

The bidimensionality theory provides strong combinatorial properties
and algorithmic results about bidimensional problems in minor-closed graph
families, unifying and improving several previous results.
The theory is based on algorithmic and combinatorial extensions to parts
of the Robertson-Seymour Graph Minor Theory, in particular initiating a
parallel theory of graph contractions.
A key combinatorial property from the theory
is that any graph in an appropriate minor-closed class has
treewidth bounded above in terms of the problem's solution value, typically by
the square root of that value.
This property leads to efficient---often subexponential---fixed-parameter
algorithms, as well as polynomial-time approximation schemes, for many
minor-closed graph classes.

The fundamental structure in the theory of bidimensionality is the
$r \times r$ grid graph.  In particular, many of the combinatorial and
algorithmic results are built upon a relation (typically linear)
between the treewidth of a graph and the size of the largest grid minor.
One such relation is known for general graphs but the bound is
superexponential:
every graph of treewidth more than $20^{2 r^5}$ has an $r \times r$ grid minor
\cite{RobSeymT94}.
This bound is usually not strong enough to derive efficient algorithms.
A substantially better, linear bound was recently established for graphs
excluding any fixed minor~$H$: every $H$-minor-free graph of treewidth
at least $c_H \, r$ has an $r \times r$ grid minor,
for some constant~$c_H$ \cite{gridminors}.
This bound generalizes similar results for smaller classes of graphs:
planar graphs \cite{mapgraphs}, bounded-genus graphs \cite{hminorfree},
and single-crossing-minor-free graphs \cite{mapgraphs,jcss}.
The bound leads to many powerful algorithmic results,
but has effectively limited those results to $H$-minor-free graphs.

In this paper we extend the bidimensionality theory to graphs that do not
exclude small minors, map graphs and power graphs,
both of which can have arbitrarily large cliques.
Given an embedded planar graph and a partition of its faces into
\emph{nations} or \emph{lakes}, the associated \emph{map graph} has a vertex
for each nation and an edge between two vertices corresponding to nations
(faces) that share a vertex.
This modified definition of the dual graph was introduced by
Chen, Grigni, and Papadimitriou \cite{ChenGP02} as a generalization of
planar graphs that can have arbitrarily large cliques.
Later Thorup \cite{Thorup98} gave a polynomial-time algorithm for
recognizing map graphs and reconstructing the planar graph and the partition.
%Thorup \cite{Thorup98} gave a polynomial-time algorithm for constructing the
%underlying embedded planar graph and face two-coloring for a given map graph,
%or determining that the given graph is not a map graph.
Recently map graphs have been studied extensively, exploiting techniques
from planar graphs, in particular in the context of
subexponential fixed-parameter algorithms and PTASs for specific
domination problems \cite{mapgraphs,Chen01}.

We can view the class of map graphs as a special case of taking powers
of a family of graphs.
The \emph{$k$th power} $G^k$ of a graph $G$ is the graph
on the same vertex set $V(G)$
with edges connecting two vertices in $G^k$
precisely if the distance between these vertices in $G$ is at most~$k$.
For a bipartite graph $G$ with bipartition $V(G) = U \cup W$,
the \emph{half-square} $G^2[U]$ is the graph on one side $U$ of the partition,
with two vertices adjacent in $G^2[U]$ precisely if the distance
between these vertices in $G$ is~$2$.
A graph is a map graph if and only if it is the half-square of some planar
bipartite graph \cite{ChenGP02}.
In fact, this translation between map graphs and half-squares is
constructive and takes polynomial time.

\subsection{Our Results and Techniques}

In this paper we establish strong (polynomial) relations between treewidth and
grid minors for map graphs and for powers of graphs.
We prove that any map graph of treewidth $r^3$ has an
$\Omega(r) \times \Omega(r)$ grid minor.
We prove that, for any graph class with a polynomial relation between
treewidth and grid minors (such as $H$-minor-free graphs and map graphs),
the family of $k$th powers of these graphs also have such a polynomial
relation, where the polynomial degree is larger by just a constant,
interestingly independent of~$k$.

These results extend bidimensionality to map graphs and power graphs,
improving the running times of a broad class of fixed-parameter algorithms
for these graphs.
Our results also build support for Robertson, Seymour, and Thomas's conjecture
that all graphs have a polynomial relation between treewidth and grid minors
\cite{RobSeymT94}.
Indeed, from our work, we refine the conjecture to state
that all graphs of treewidth $\Omega(r^3)$ have an
$\Omega(r) \times \Omega(r)$ grid minor, and that this bound is tight.
The previous best treewidth-grid relations for map graphs and power graphs
was the superexponential bound from \cite{RobSeymT94}.

The main technique in this paper is to use approximate max-min relations
between the size of a grid minor and treewidth.  In contrast,
most previous work uses the seminal approximate max-min relation between
tangles and treewidth, or the max-min relation between tangles and branchwidth,
proved by Robertson and Seymour \cite{RobertsonS10}.
We show that grids are powerful structures that are easy to work with.
By bootstrapping, we use grids and their connections to treewidth
even to prove relations between grids and treewidth.

Another example of the power of our technique is a result we obtain
as a byproduct of our study of map graphs: every bounded-genus graph
has treewidth within a constant factor of the treewidth of its dual.
This result generalizes a conjecture of Seymour and Thomas
\cite{ST94} that the treewidth of a planar graph is within an
additive $1$ of the treewidth of its dual, which has apparently been proved
in \cite{Lapoire,BMT01} using a complicated approach.
Such a primal-dual treewidth relation is useful e.g.\ for bounding the change
in treewidth when performing operations in the dual.
In the case of our result, we can bound the change in treewidth of a
bounded-genus graph when manipulating faces, e.g., when contracting a face
down to a point as in \cite{gridminors}.
Our proof crucially uses the connections between treewidth and
grid minors, and this approach leads to a relatively clean argument.
The tools we use come from bidimensionality theory and graph contractions,
even though the result is not explicitly about either.

\subsection{Algorithmic and Combinatorial Applications}

Our treewidth-grid relations have several useful consequences
with respect to fixed-parameter algorithms, minor-bidimensionality,
and parameter-treewidth bounds.

A \emph{fixed-parameter algorithm} is an algorithm for computing a parameter
$P(G)$ of a graph $G$ whose running time is $h(P(G)) \, n^{O(1)}$
for some function~$h$.  A typical function $h$ for many fixed-parameter
algorithms is $h(k) = 2^{O(k)}$.
A celebrated example of a fixed-parameter-tractable problem is
vertex cover, asking whether an input graph has at most $k$ vertices
that are incident to all its edges, which admits a solution as fast as
$O(k n + 1.285^k)$~\cite{CKJ99}.
For more results about fixed-parameter tractability and intractability,
see the book of Downey and Fellows \cite{DowneyF99}.

A major recent approach for obtaining efficient fixed-parameter algorithms
is through ``parameter-treewidth bounds'',
a notion at the heart of bidimensionality.
A \emph{parameter-treewidth bound} is an upper bound $f(k)$ on the treewidth
of a graph with parameter value~$k$.
Typically, $f(k)$ is polynomial in~$k$.
%In many cases, $f(k)$ can even be shown to be sublinear in~$k$,
%often $O(\sqrt k)$.
Parameter-treewidth bounds have been established for many parameters;
see, e.g.,
\cite{AlberBFKN02, KanjPer02, FominT02, AFN01, CKL01, FVS01, GKL01,
mapgraphs, jcss, APPROX, DHT02, localtw, apexfree, hminorfree}.
Essentially all of these bounds can be obtained from the general theory
of bidimensional parameters (see, e.g., \cite{gdsurvey}).
Thus bidimensionality is the most powerful method so far
for establishing parameter-treewidth bounds,
encompassing all such previous results for $H$-minor-free graphs.
However, all of these results are limited to graphs that exclude a fixed minor.

A parameter is \emph{minor-bidimensional}
if it is at least $g(r)$ in the $r \times r$ grid graph
and if the parameter does not increase when taking minors.
Examples of minor-bidimensional parameters
include the number of vertices and the size of various structures,
e.g., feedback vertex set, vertex cover, minimum maximal matching,
face cover, and a series of vertex-removal parameters.
Tight parameter-treewidth bounds have been established for all
minor-bidimensional parameters in $H$-minor-free graphs for any fixed graph~$H$
\cite{gridminors,localtw,hminorfree}.

Our results provide polynomial parameter-treewidth bounds for all
minor-bidimensional parameters in map graphs and power graphs:

\begin{theorem} \label{parameter-treewidth bound}
  For any minor-bidimensional parameter~$P$
  which is at least $g(r)$ in the $r \times r$ grid,
  every map graph $G$ has treewidth $\tw(G) = O(g^{-1}(P(G)))^3$.
  More generally suppose that, if graph $G$ has treewidth at least $c r^\alpha$
  for constants $c, \alpha > 0$, then $G$ has an $r \times r$ grid minor.
  Then, for any even (respectively, odd) integer $k \geq 1$,
  $G^k$ has treewidth $\tw(G) = O(g^{-1}(P(G)))^{\alpha+4}$
  (respectively, $\tw(G) = O(g^{-1}(P(G)))^{\alpha+6}$).
  In particular, for $H$-minor-free graphs~$G$,
  and for any even (respectively, odd) integer $k \geq 1$,
  $G^k$ has treewidth $\tw(G) = O(g^{-1}(P(G)))^5$
  (respectively, $\tw(G) = O(g^{-1}(P(G)))^7$).
\end{theorem}

This result naturally leads to a collection of fixed-parameter algorithms,
using commonly available algorithms for graphs of bounded treewidth:

\begin{corollary}
  Consider a parameter $P$ that can be computed on a graph $G$
  in $h(w) \, n^{O(1)}$ time
  given a tree decomposition of $G$ of width at most~$w$.
  If $P$ is minor-bidimensional and
  at least $g(r)$ in the $r \times r$ grid,
  then there is an algorithm computing $P$ on any map graph or power graph $G$
  with running time $[h(O(g^{-1}(k))^\beta) + 2^{O(g^{-1}(k))^\beta}]
  \, n^{O(1)}$, where $\beta$ is the degree of $O(g^{-1}(P(G))$ in the
  polynomial treewidth bound from Theorem~\ref{parameter-treewidth bound}.
  In particular, if $h(w) = 2^{O(w)}$ and $g(k) = \Omega(k^2)$,
  then the running time is $2^{O(k^{\beta/2})} n^{O(1)}$.
\end{corollary}

The proofs of these consequences follow directly from combining
\cite{localtw} with Theorems \ref{map grid} and \ref{power grid} below.

In contrast, the best previous results for this general family of problems
in these graph families have running times
$[h(2^{O(g^{-1}(k))^5}) + 2^{2^{O(g^{-1}(k))^5}}] \, n^{O(1)}$
\cite{localtw,generalgraphs}.

\iffalse

. general graphs FPT

(mention \cite{firstorder} and maybe \cite{mapgraphs})

. parameter-treewidth bounds (for minor-bidimensional)

. better FPT (for minor-bidimensional)

. bidimensionality

lower bound $r^2 lg r$

. foundation: grid minors

. grid minors

general graphs RS

theorems:

\begin{theorem}[\cite{RobSeymT94}] \label{planar grid}
  Every planar graph of treewidth $w$ has an
  $\Omega(w+1) \times \Omega(w+1)$ grid graph as a minor.%
  %
  \footnote{We require bounds involving asymptotic notation
    $O$, $\Omega$, and $\Theta$ to hold for all values of the parameters,
    in particular,~$w$.  Thus, $\Omega(w+1)$ has a different meaning
    from $\Omega(w)$ when $w = 0$.
    In this theorem, when the treewidth is $0$, i.e., the graph has no edges,
    there is still a $1 \times 1$ grid.
  }
\end{theorem}

Constant in this theorem is small.

\begin{theorem}[\cite{gridminors}] \label{H-minor-free grid}
  For any fixed graph $H$,
  every $H$-minor-free graph of treewidth $w$ has an
  $\Omega(w+1) \times \Omega(w+1)$ grid graph as a minor.
\end{theorem}

. our results (combinatorial)

map graphs

power graphs

tightening results for general graphs (how big a grid)

our technique: use grids to deal with treewidth (instead of tangles)

application of our technique, and bidimensionality theory:
treewidth relation to dual (connection to radial)

Algorithmic Applications:

. general graphs FPT

(mention \cite{firstorder} and maybe \cite{mapgraphs})

. parameter-treewidth bounds (for minor-bidimensional)

. better FPT (for minor-bidimensional)

\fi

\section{Definitions and Preliminaries}

\paragraph{Treewidth.}

The notion of treewidth was introduced by Robertson and Seymour
\cite{Robert86}. To define this notion, first we consider a
representation of a graph as a tree, called a tree decomposition.
Precisely, a \emph{tree decomposition} of a graph $G=(V, E)$
is a pair $(T, \chi)$ in which $T=(I, F)$ is a tree and
$\chi=\{\chi_i \mid i\in I\}$ is a family of subsets of $V(G)$ such that
\begin{enumerate}
\item $\bigcup_{i\in I}\chi_i= V$;
\item for each edge $e=\{u, v\} \in E$,
  there exists an $i\in I$ such that both $u$ and $v$
  belong to $\chi_i$; and
\item for all $v\in V$, the set of nodes
  $\{i\in I \mid v \in \chi_i\}$ forms a connected subtree of~$T$.
\end{enumerate}
To distinguish between vertices of the original graph $G$ and vertices
of $T$ in the tree decomposition, we call vertices of $T$ {\em nodes} and their
corresponding $\chi_i$'s {\em bags}.
The {\em width} of the tree decomposition is the maximum size of a bag in
$\chi$ minus~$1$.
The {\em treewidth} of a graph $G$, denoted $\tw(G)$, is the
minimum width over all possible tree decompositions of $G$.
%A tree decomposition is called a {\em path decomposition} if  $T=(I, F)$
%is a path.  The {\em pathwidth} of a graph $G$, denoted $\pw(G)$, is the
%minimum width over all possible path decompositions of~$G$.

\paragraph{Minors and contractions.}
Given an edge $e = \{v,w\}$ in a graph $G$, the \emph{contraction} of $e$
in $G$ is the result of identifying vertices $v$ and $w$ in $G$
and removing all loops and duplicate edges.
A graph $H$ obtained by a sequence of such edge contractions starting from $G$
is said to be a \emph{contraction} of~$G$.
A graph $H$ is a \emph{minor} of $G$ if $H$ is a subgraph of some
contraction of~$G$.
A graph class ${\cal C}$ is \emph{minor-closed} if any minor of
any graph in ${\cal C}$ is also a member of ${\cal C}$.
A minor-closed graph class ${\cal C}$ is \emph{$H$-minor-free}
if $H \notin {\cal C}$.  More generally, we use the term ``$H$-minor-free''
to refer to any minor-closed graph class that excludes some fixed graph~$H$.

\paragraph{Grid minors.}
We use the following important connections between treewidth and the
size of the largest grid minor.
The \emph{$r \times r$ grid} is the planar graph with $r^2$ vertices
arranged on a square grid and with edges connecting horizontally and
vertically adjacent vertices.
First we state the connection for planar graphs:

\begin{theorem}[\cite{RobSeymT94}] \label{planar grid}
  Every planar graph of treewidth $w$ has an
  $\Omega(w+1) \times \Omega(w+1)$ grid graph as a minor.%
  \footnote{We require bounds involving asymptotic notation
    $O$, $\Omega$, and $\Theta$ to hold for all values of the parameters,
    in particular,~$w$.  Thus, $\Omega(w+1)$ has a different meaning
    from $\Omega(w)$ when $w = 0$.
    In this theorem, when the treewidth is $0$, i.e., the graph has no edges,
    there is still a $1 \times 1$ grid.
  }
\end{theorem}

The more general connection for $H$-minor-free graphs
has been obtained recently:

\begin{theorem}[\cite{gridminors}] \label{H-minor-free grid}
  For any fixed graph $H$,
  every $H$-minor-free graph of treewidth $w$ has an
  $\Omega(w+1) \times \Omega(w+1)$ grid graph as a minor.
\end{theorem}

\paragraph{Embeddings.}
A \emph{2-cell embedding} of a graph $G$ in a surface $\Sigma$
(two-dimensional manifold)
is a drawing of the vertices as points in $\Sigma$
and the edges as curves in $\Sigma$ such that
no two points coincide, two curves intersect only at shared endpoints,
and every face (region) bounded by edges is an open disk.
We define the \emph{Euler genus} or simply \emph{genus} of a surface $\Sigma$
to be the ``non-orientable genus'' or ``crosscap number'' for non-orientable
surfaces $\Sigma$, and twice the ``orientable genus'' or ``handle number''
for orientable surfaces $\Sigma$.
The \emph{(Euler) genus} of a graph $G$ is the minimum genus of a surface
in which $G$ can be 2-cell embedded.
A graph has \emph{bounded genus} if its genus is $O(1)$.

A \emph{planar embedding} is a 2-cell embedding into the
plane (topological sphere).
An \emph{embedded planar graph} is a graph together with a planar embedding.

\iffalse
Combinatorially, \emph{contracting} an edge $e = \{u,v\}$ in a graph
is the operation of removing the edge $e$ and then
replacing the two vertices $u$ and $v$ (and all their occurrences as endpoints
of other edges) by a single (new) vertex~$w$.
%$G' = (V - \{u,v\} \cup \{w\}, )$.
This definition may produce multiple edges and loops.
%whose neighbors are all vertices that were neighbors of $u$ or
%$v$ (removing any duplicates), except $u$ and $v$ themselves.
%This operation implicitly removes multiple edges resulting from
%unifying $u$ and~$v$.
Topologically, \emph{contracting} an edge $e = \{u,v\}$ in an
embedded planar graph is an operation that produces the same graph
as the combinatorial sense of contraction
but also ``preserves'' the planar embedding.
More precisely, we continuously move $u$ along the path along which the edge
$e$ is embedded, dragging the edges incident to $u$ along with $u$,
other than the edge $e$ which simply gets shorter, until
$u$ coincides with~$v$ and the edge $e$ disappears.
Again this definition may produce multiple edges and loops.
\fi

\paragraph{Map graphs.}
We define a map graph and related notions in terms of an
embedded planar graph $G$ and a partition of faces
into a collection $N(G)$ of \emph{nations} and a collection $L(G)$
of \emph{lakes}.  Thus, $N(G) \cup L(G)$ is the set of faces of~$G$.

We define the \emph{(modified) dual} $D = D(G)$ of $G$ in terms of
only the nations of~$G$.
$D$ has a vertex for every nation of~$G$,
and two vertices are adjacent in $D$ if the corresponding nations of $G$
share an edge.

The \emph{map graph} $M = M(G)$ of $G$ has
a vertex for every nation of $G$, and two vertices are adjacent in $M(G)$
if the corresponding nations of $G$ share a vertex.
The map graph $M(G)$ is a subgraph of the dual graph~$D(G)$.

\paragraph{Canonical map graphs.}

We canonicalize $G$ in the following ways that preserve the map graph $M(G)$.
First, we remove any vertex of $G$ incident only to lakes,
because it and its incident edges do not contribute to the map graph $M(G)$.
%Second, we repeatedly contract any degree-two vertex into one of its neighbors.
Second, for any edge of $G$ whose two incident faces are both lakes
(possibly the same lake),
we delete the edge and merge the corresponding lakes,
because again this will not change the map graph~$M(G)$.

Third, we modify $G$ to ensure that every vertex is incident to at most
one lake, and incident to such a lake at most once.
Consider a vertex $v$ that violates this property, and suppose there is an
incident lake between edges $\{v, w_i\}$ and $\{v, w'_i\}$
for $i = 1, 2, \dots, l$.
We split $v$ into $l+1$ vertices $v, v_1, v_2, \dots, v_l$,
with $v_i$ placed near $v$ in the wedge $w_i, v, w'_i\}$.
We connect these $l+1$ vertices in a star,
with an edge between $v$ and $v_i$ for $i = 1, 2, \dots, l$.
Edges $\{v, w_i\}$ and $\{v, w'_i\}$ reroute to be
$\{v_i, w_i\}$ and $\{v_i, w'_i\}$,
and all other edges incident to $v$ remain as they were.
%Finally, if $l=2$, then we contract $v$ into either $v_1$ or $v_2$,
as in the second canonicalization.
This modification preserves the map graph $M(G)$
and results in no lakes touching at~$v$.

Finally, we assume that the map graph $M(G)$ is connected, i.e.,
a lake never separates two nations in~$G$,
because we can always consider each connected component separately.

\paragraph{Radial graphs.}
The \emph{radial graph} $R = R(G)$ has a vertex for every vertex of $G$
and for every nation of~$G$, and we label them the same:
$V(R) = V(G) \cup N(G)$.
$R(G)$ is bipartite with this bipartition.
Two vertices $v \in V(G)$ and $f \in N(G)$ are adjacent in $R(G)$
if their corresponding vertex $v$ and nation $f$ are incident.

We also consider the union graph $R \cup D$.
$R \cup D$ has the same vertex set as the radial graph~$R$,
which is a superset of the vertex set of the dual graph~$D$.
The edges in $R \cup D$ consist of all edges in $R$ and all edges in~$D$.

We also define the \emph{radial graph} $R = R(G)$ for a graph $G$
2-cell embedded in an arbitrary surface~$\Sigma$.  In this case,
we do not allow lakes, and consider every face to be a nation.
Otherwise, the definition is the same.

\section{Treewidth-Grid Relation for Map Graphs}
\label{map graphs}

In this section we prove a polynomial relation between the treewidth of
a map graph and the size of the largest grid minor.
The main idea is to relate the treewidth of the map graph $M(G)$,
the treewidth of the radial graph $R(G)$,
the treewidth of the dual graph $D(G)$,
and the treewidth of the union graph $R(G) \cup D(G)$.

\xxx{Actually, in the following lemma,
  we really only need the direction that we prove at length,
  and we don't need the other direction, and therefore we can use $R$
  directly instead of $R \cup G$.}

\begin{lemma} \label{radial tw}
  The treewidth of the union $R \cup D$ of the radial graph $R$
  and the dual graph $D$, plus~$1$,
  is within a constant factor of the treewidth of the dual graph~$D$,
  plus~$1$.
\end{lemma}

\begin{proof}
  First, $\tw(D) + 1 \leq \tw(R \cup D) + 1$
  because $D$ is a subgraph of $R \cup D$.

  The rest of the proof establishes that $\tw(D)+1 = \Omega(\tw(R \cup D)+1)$.
  Because both graphs are planar, we know by Theorem \ref{planar grid}
  that $1$ plus the treewidth of either graph is within a constant factor
  of the dimension of the largest grid minor.
  Thus it suffices to show that we can convert
  a given $k \times k$ grid minor $K$ of $R \cup D$
  into an $\Omega(k) \times \Omega(k)$ grid minor of~$D$.

  Consider the sequence of edge contractions and removals that bring
  $R \cup D$ to the grid~$K$.  Discard all edge deletions from this
  sequence, but remove any loops and duplicate copies of edges that
  arise from contractions.
  The resulting graph $K'$ remains planar and has the same vertices as~$K$,
  and therefore $K'$ is a partially triangulated $k \times k$ grid,
  in the sense that each face of the $k \times k$ grid
  can have a noncrossing set of additional edges.
  (All bounded faces of the grid have $4$ vertices
  and so at most one additional edge.)

  We label each vertex $v$ in $K'$ with the set of vertices
  from $R \cup D$ that contracted to form~$v$.
  We call $v$ \emph{facial} if at least one of these vertices
  is a vertex of the dual graph~$D$.
  Otherwise, $v$ is \emph{nonfacial}.
  No two nonfacial vertices can be adjacent in~$K'$, because
  no two vertices in $G$ are adjacent in $R \cup D$.

  Assign coordinates $(x,y)$, $0 \leq x,y < k$, to each vertex $v$ in~$K'$.
  We assume without loss of generality that $k$ is divisible by~$6$
  (decreasing $k$ by at most $5$ if necessary).
  For each $i, j$ with $1 \leq i,j \leq k/6 - 1$,
  either vertex $(6i+1,6j+1)$ or vertex $(6i+2,6j+1)$
  is facial, because these two vertices are adjacent in~$K'$.
  Let $v_{i,j}$ denote a facial vertex among this pair.
  Let $\hat v_{i,j}$ denote a vertex of the dual graph $D$
  in the label of $v_{i,j}$ (which exists by the definition of facial).

  For any $i, j$ with $1 \leq i \leq k/6 - 1$ and $1 \leq j \leq k/6 - 2$,
  we claim that there is a simple path
  between $\hat v_{i,j}$ and $\hat v_{i,j+1}$ in~$D$
  using only vertices in $D$ that appear in the labels of
  vertices in $R'$ with coordinates in
  the rectangle $(6i \twodots 6i+3, 6j \twodots 6(j+1)+3)$.
  We start with a shortest path $P_{K'}$
  between $v_{i,j}$ and $v_{i,j+1}$ in~$K'$,
  which is simple and
  remains in the subrectangle $(6i+1 \twodots 6i+2, 6j+1 \twodots 6(j+1)+2)$.
  We convert $P_{K'}$ into a simple path $P_{R \cup D}$
  between $\hat v_{i,j}$ and $\hat v_{i,j+1}$ in $R \cup D$ using
  only the vertices in $R \cup D$ that appear in the labels of the vertices
  in $K'$ along~$P_{K'}$.
  Here we use that the subgraph of $R \cup D$ induced by the label set of a
  vertex in $K'$ is connected, because that vertex in $K'$ was formed by
  contracting edges in this subgraph.
  For each edge in the path~$P_{K'}$, we pick an edge in $R \cup D$ that
  forms it as a result of the contractions;
  then we connect together the endpoints of these
  edges, and connect the first and last edges to $\hat v_{i,j}$ and
  $\hat v_{i,j+1}$ respectively, by finding shortest paths within the
  subgraphs of $R \cup D$ induced by label sets.
  Finally we convert this path $P_{R \cup D}$
  into a simple path $P_D$ in $D$
  with the desired properties.
  The vertices along the path $P_{R \cup D}$ divide into two classes:
  those in $D$ (corresponding to nations of~$G$) and those in $G$
  (corresponding to vertices of~$G$).
  Among the subsequence of vertices along the path $P_{R \cup D}$,
  restricted to vertices in~$D$,
  we claim that every two consecutive vertices $v,w$ can be connected
  using only vertices in $D$
  that appear in the labels of vertices in the desired rectangle.
  If $v$ and $w$ are consecutive along the path $P_{R \cup D}$,
  then they are adjacent in $D$ and we are done.
  Otherwise, $v$ and $w$ are separated in the path $P_{R \cup D}$
  by one vertex $u$ of $G$ (because no two vertices of $G$ are
  adjacent in $R \cup D$).
  In~$G$, this situation corresponds to two nations $v$ and $w$
  that share the vertex~$u$.
  Because of our canonicalization, $u$ is incident to at most one lake,
  at most once, and therefore there is a sequence of nations
  $v = f_1, f_2, \dots, f_j = w$ in clockwise or counterclockwise
  order around~$u$.  Thus in $D$ we obtain a path
  $v = f_1, f_2, \dots, f_j = w$.
  Each $f_i$ is incident to $u$ and therefore has distance $1$ from $u$
  in $R \cup D$.
  Because the contractions that formed $K'$ from $R \cup D$
  only decrease distances, the vertices of $K'$ with labels
  including $f_i$ and $u$ have distance at most $1$ in~$K'$.
  Therefore each $f_i$ is in a label of a vertex within the thickened rectangle
  $(6i \twodots 6i+3, 6j \twodots 6(j+1)+3)$.
  If the path is not simple, we can take the shortest path
  between its endpoints in the subgraph induced by the vertices of the path,
  and obtain a simple path.

  Symmetrically,
  for any $i, j$ with $1 \leq i \leq k/6 - 2$ and $1 \leq j \leq k/6 - 1$,
  we obtain that there is a simple path
  between $\hat v_{i,j}$ and $\hat v_{i+1,j}$ in~$D$
  using only vertices in $D$ that appear in the labels of
  vertices in $R'$ with coordinates in
  the rectangle $(6i \twodots 6(i+1)+3, 6j \twodots 6j+3)$.

  We construct a grid minor $K''$ of $D$ as follows.
  We start with the union, over all $i,j$,
  of the simple path between $\hat v_{i,j}$ and $\hat v_{i,j+1}$ in~$D$
  and the simple path between $\hat v_{i,j}$ and $\hat v_{i+1,j}$ in~$D$.
  (In other words, we delete all vertices not belonging to one of these paths.)
  Then we contract every vertex in this union
  that is not one of the $\hat v_{i,j}$'s
  toward its ``nearest'' $\hat v_{i,j}$.
  More precisely, for each path between $\hat v_{i,j}$ and $\hat v_{i,j+1}$,
  we cut the path at the first edge that crosses from
  row $6i+4$ to row $6i+5$; then we contract all vertices in the path
  before the cut into $\hat v_{i,j}$, and we contract all vertices in the path
  after the cut into $\hat v_{i,j+1}$.
  Similarly we cut each path between $\hat v_{i,j}$ and $\hat v_{i+1,j}$
  at the first edge that crosses from column $6i+4$ to column $6i+5$,
  and contract accordingly.
  Because of the rectangular bounds on each path,
  the rectangle $(6i \twodots 6i+3, 6j+4 \twodots 6j+5)$
  is intersected by a unique path,
  the one from $\hat v_{i,j}$ to~$\hat v_{i,j+1}$,
  and the rectangle $(6i+4 \twodots 6i+5, 6j \twodots 6j+3)$
  is intersected by a unique path,
  the one from $\hat v_{i,j}$ to~$\hat v_{i+1,j}$.
  Hence our contraction process does not merge paths that
  were not originally incident (at one of the~$\hat v_{i,j}$'s).
  Also, because each path is simple and strays by distance at most $1$
  from the original shortest path in the grid~$K'$, the vertices before
  the cut are disjoint from the vertices after the cut in the path.
  Therefore, each vertex on a path contracts to a unique vertex~$\hat v_{i,j}$,
  and each path contracts to a single edge between $\hat v_{i,j}$ and
  either $\hat v_{i,j+1}$ or~$\hat v_{i+1,j}$.
  Thus we obtain a $(k/6-1) \times (k/6-1)$ grid minor $K''$ of~$D$.
\end{proof}

\begin{lemma} \label{map tw}
  The treewidth of the map graph $M$ is at most
  the product of the maximum degree of a vertex in $G$
  and $\tw(R) + 1$, one more than the treewidth of the radial graph~$R$.
\end{lemma}

\begin{proof}
  Suppose we have a tree decomposition $(T,\chi)$ of the radial graph $R$
  of width~$w$.
  We modify this tree decomposition into another tree decomposition
  $(T,\chi')$ by replacing each occurrence of a vertex $v \in V(G)$
  in a bag $\bag$ of $\chi$ with all nations incident to~$v$.
  Thus, bags in $\chi'$ consist only of nations.

  We claim $(T,\chi')$ is a tree decomposition of~$M$.
  First, observe that every vertex of the map graph $M$
  appears in some bag $\bag$ of $\chi'$, because nations
  are vertices in the radial graph as well, so every nation
  appears in a bag of~$\chi$.

  Second, we claim that every vertex of the map graph $M$
  appears in a connected subtree of bags in $(T,\chi')$.
  A nation $f$ appears in a bag $\bag'$ of $\chi'$
  if either it appears in the corresponding bag $\bag$ of $\chi$
  or one of its vertices appears in corresponding bag $\bag$ of $\chi$.
  The set of bags in $\chi$ containing the nation $f$
  forms a connected subtree of $T$,
  and the set of bags in $\chi$ containing any vertex $v$ of $f$
  forms a connected subtree of~$T$.
  These two subtrees, for any choice of $v$, overlap in at least one
  node of $T$ because $v$ and $f$ are adjacent in the radial graph~$R$,
  and thus this edge $(v,f)$ appeared in some bag of~$\chi$.
  Therefore the union of the subtree of $T$ induced by $f$
  and all vertices $v$ of $f$ is connected.
  This union is precisely the set of nodes in $T$ whose bags in $\chi'$
  contain~$f$.

  Third, we claim that every edge of the map graph $M$
  appears in some bag of $\chi'$.
  An edge arises in $M$ when two nations $f_1, f_2$ share a vertex $v$ in~$G$.
  This vertex $v$ appears in some bag $\bag$ of $\chi$,
  and in constructing $\chi'$ we replaced $v$ with nations $f_1$, $f_2$,
  and possibly other nations.  Therefore $f_1$ and $f_2$ appear
  in the corresponding bag $\bag'$ of $\chi'$.

  Finally we claim that the size of any bag $\bag'$ in $\chi'$ is at most 
  the maximum degree $\Delta$ of a vertex in $G$ times the size of the
  corresponding bag $\bag$ in~$\chi$.  This claim follows from the
  construction because each vertex is replaced by at most $\Delta$
  nations in the transformation from $\bag$ to~$\bag'$.
  The size of each original bag $\bag$ in $\chi$ is at most
  one more than the treewidth of $R$.
  Therefore the maximum bag size in $\chi'$ is at most
  $\Delta (\tw(R)+1)$, and the treewidth of $M$ is at most one less
  than this maximum bag size.
\end{proof}

\begin{theorem} \label{map grid}
  If the treewidth of the map graph $M$ is $r^3$,
  then it has an $\Omega(r) \times \Omega(r)$ grid as a minor.
\end{theorem}

\begin{proof}
  By Lemma \ref{map tw}, $\tw(M) = O(\Delta \cdot \tw(R))$.
  Because $R$ is a subgraph of $R \cup D$,
  $\tw(M) = O(\Delta \cdot \tw(R \cup D))$.
  By Lemma \ref{radial tw}, $\tw(M) = O(\Delta \cdot (\tw(D)+1))$.
  Thus, if $\tw(M) = \Omega(r^3)$, then
  either $\tw(D) = \Omega(r)$ or $\Delta = \Omega(r^2)$.
  In the former case, $D$ is a planar subgraph of $M$
  and therefore $D$ and $M$ have an $\Omega(r) \times \Omega(r)$ grid
  as a minor by Theorem \ref{planar grid}.
  In the latter case, $M$ has a $K_\Delta = K_{\Omega(r^2)}$ clique
  as a subgraph, and therefore has an $\Omega(r) \times \Omega(r)$
  grid as minor.
\end{proof}

Next we show that this theorem cannot be improved from
$\Omega(r^3)$ to anything~$o(r^2)$:

\begin{proposition}
  There are map graphs whose treewidth is $r^2-1$
  and whose largest grid minor is $r \times r$.
\end{proposition}

\begin{proof}
  Let $G$ be an embedded wheel graph with $r^2$ spokes.
  We set all $r^2$ bounded faces to be nations
  and the exterior face to be a lake.
  Then the dual graph $D$ is a cycle,
  and the map graph $M$ is the clique~$K_{r^2}$.
  Therefore $M$ has treewidth~$r^2-1$,
  yet its smallest grid minor is $r \times r$.
\end{proof}

Robertson, Seymour, and Thomas \cite{RobSeymT94} prove a stronger lower bound
of $\Theta(r^2 \lg r)$ but only for the case of general graphs.

\xxx{$r^3$ lower bound?  Idea:
  Take the wheel construction with $r^2$ spokes,
  divide the faces into four equal groups,
  squash the wheel into a square,
  and then glue them together edge-to-edge into an $r \times r$ grid.
  Claim 1: Largest grid minor is $r \times r$
  (either from a subset of the cliques $K_{r^2}$ or from the
  overall $r \times r$ grid).
  [Actually this may not be true with this construction...]
  Claim 2: Treewidth is $\Theta(r^3)$
  (a bit tricky).

  This would improve best lower bound for arbitrary graphs.}

\section{Treewidth-Grid Relation for Power Graphs}

In this section we prove a polynomial relation between the treewidth of
a power graph and the size of the largest grid minor.
The technique here is quite different, analyzing how a radius-$r$ neighborhood
in the graph can be covered by radius-$(r/2)$ neighborhoods---a kind of
``sphere packing'' argument.

\begin{theorem} \label{power grid}
  Suppose that, if graph $G$ has treewidth at least $c r^\alpha$
  for constants $c, \alpha > 0$, then $G$ has an $r \times r$ grid minor.
  For any even (respectively, odd) integer $k \geq 1$,
  if $G^k$ has treewidth at least
  $c r^{\alpha+4}$ (respectively, $c r^{\alpha+6}$),
  then it has an $r \times r$ grid minor.
\end{theorem}

\begin{proof}
  Let $\maxdeg(G^k)$ denote the maximum degree of any vertex in~$G^k$,
  that is, the maximum size of the $k$-neighborhood of a vertex in~$G$.
  First we claim that $\tw(G^k) \leq \maxdeg(G^k) \tw(G)$.
  Consider a tree decomposition $(T,\chi)$ of~$G$.
  Replace each occurrence of vertex $v$ in $\chi_x$
  with the entire radius-$k$ neighborhood of $v$ in~$G$.
  Thus we expand the maximum bag size by a factor of at most $\maxdeg(G^k)$,
  and the width of the resulting $(T,\chi')$ is at most
  $\maxdeg(G^k) (\tw(G) + 1)$.
  We claim that $(T,\chi')$ is a tree decomposition of~$G^k$.
  First, if two vertices $v$ and $w$ are adjacent in~$G^k$, i.e.,
  within distance $k$ in~$G$, then by construction they are in a common bag
  in $(T,\chi')$, indeed any bag that originally contained either $v$ or~$w$.
  Second, we claim that the set of bags containing a vertex $v$ is a connected
  subtree of~$T$.
  In other words, we claim that any two vertices $u$ and $w$ that are within
  distance $k$ of $v$, which give rise to occurrences of~$v$ in $\chi'$,
  can be connected via a path in $T$ along which the bags always contain~$v$.
  Concatenate the shortest path $u = v_0, v_1, \dots, v_j = v$ from $u$ to $v$
  in $G$ and the shortest path $v = v_j, v_{j+1}, \dots, v_l = w$ from $v$ to
  $w$ in $G$, both of which use vertices $v_i$ always within distance
  $k$ of~$v$.
  Now construct the desired path in $T$ by visiting, for each $i$ in turn,
  the subtree of bags in $\chi$ containing occurrences of~$v_i$,
  whose corresponding bags in $\chi'$ contain occurrences of~$v$.
  Here we use that the bags in $\chi$ containing occurrences of $v_i$
  form a connected subtree of~$T$, and that this subtree for $v_i$
  and this subtree for $v_{i+1}$ share a node because $v_i$ is adjacent
  to~$v_{i+1}$.

  If $\tw(G^k) \geq c r^{\alpha+4}$,
  then either $\maxdeg(G^k) \geq r^4$ or $\tw(G) \geq c r^\alpha$.
  In the latter case, we obtain by supposition that
  $G$ has an $r \times r$ grid minor and thus so does the supergraph~$G^k$.
  Therefore we concentrate on the former case when
  $\maxdeg(G^k) \geq r^4$.
  Let $v$ be the vertex in $G$ whose $k$-neighborhood $N_k$ has maximum size,
  $\Delta(G^k)$.
  There are two cases depending on whether $k$ is even or odd.

  The simpler case is when $k$ is even.
  If the $(k/2)$-neighborhood $N_{k/2}$ of $v$ in $G$ has size at least~$r^2$,
  then in $G^k$ we obtain a clique $K_{r^2}$ on those vertices,
  so we obtain an $r \times r$ grid minor.
  Otherwise, label each vertex in the $k$-neighborhood $N_k$ with the nearest
  vertex in the $(k/2)$-neighborhood~$N_{k/2}$.
  If any vertex in the $(k/2)$-neighborhood $N_{k/2}$ is assigned as the
  label to at least $r^2$ vertices in~$N_k$,
  then again we obtain a $K_{r^2}$ clique subgraph in $G^k$
  and thus an $r \times r$ grid minor.
  Otherwise, the $k$-neighborhood $N_k$ has size
  strictly less than $r^2 \cdot r^2 = r^4$,
  contradicting that $|N_k| = \maxdeg(G^k) \geq r^4$.

  The case when $k$ is odd is similar.
  As before, if the $\lfloor k/2 \rfloor$-neighborhood
  $N_{\lfloor k/2 \rfloor}$ of $v$ in $G$ has size at least~$r^2$, then
  in $G^k$ we obtain a clique $K_{r^2}$ and thus an $r \times r$ grid minor.
  Otherwise, label each vertex in the $(k-1)$-neighborhood $N_{k-1}$ of $v$
  with the nearest vertex in the $\lfloor k/2 \rfloor$-neighborhood
  $N_{\lfloor k/2 \rfloor}$.
  If any vertex in the $\lfloor k/2 \rfloor$-neighborhood
  $N_{\lfloor k/2 \rfloor}$ is assigned as the label to at least $r^2$
  vertices in~$N_{k-1}$, then again we obtain a $K_{r^2}$ clique
  and an $r \times r$ grid.
  Otherwise, $|N_{k-1}| < r^4$.
  Finally label each vertex in $N_k$ with the nearest vertex in~$N_{k-1}$.
  If any vertex in $N_{k-1}$ is assigned as the label to at least $r^2$
  vertices in~$N_k$, then again we obtain a $K_{r^2}$ clique
  and an $r \times r$ grid.
  Otherwise, $|N_k| < r^4 \cdot r^2 = r^6$,
  contradicting that $|N_k| = \maxdeg(G^k) \geq r^6$.
\end{proof}

We have the following immediate consequence of
Theorems \ref{H-minor-free grid}, \ref{map grid}, and~\ref{power grid}:

\begin{corollary} \label{H-minor-free power grid}
  For any $H$-minor-free graph~$G$,
  and for any even (respectively, odd) integer $k \geq 1$,
  if $G^k$ has treewidth at least~$r^5$ (respectively,~$r^7$),
  then it has an $\Omega(r) \times \Omega(r)$ grid minor.
  For any map graph~$G$,
  and for any even (respectively, odd) integer $k \geq 1$,
  if $G^k$ has treewidth at least~$r^7$ (respectively,~$r^9$),
  then it has an $\Omega(r) \times \Omega(r)$ grid minor.
\end{corollary}

\section{Primal-Dual Treewidth Relation for Bounded-Genus Graphs}

Robertson and Seymour \cite{RobertsonS-XI,ST94}
proved that the branchwidth of a planar
graph is equal to the branchwidth of its dual,
and conjectured that the treewidth of a planar graph
is within an additive $1$ of the treewidth of its dual.
The latter conjecture was apparently proved in \cite{Lapoire,BMT01},
though the proof is complicated.
Here we prove that the treewidth (and hence the branchwidth) of any
graph 2-cell embedded in a bounded-genus surface is within a constant
factor of the treewidth of its dual.
Thus the result applies more generally, though the connection is slightly
weaker (constant factor instead of additive constant).

We crucially use the connection between treewidth and grids to obtain a
relatively simple proof of this result.
Our proof uses Section \ref{map graphs}, generalized to the
bounded-genus case, and forbidding lakes.

We need the following theorem from the contraction bidimensionality theory,
and a simple corollary.

\begin{theorem}[\cite{boundedgenus}] \label{handle}
  %For any $k \geq 12 \eg(G)$,
  %%if $G$ has banchwidth more than $4 k (\eg(G)+1)$,
  %%i.e. tw(G)+1 > 6 k (\eg(G)+1)  [because tw+1 \leq 3/2 bw]
  %%i.e. tw(G) \geq 6 k (\eg(G)+1)
  %if $G$ has treewidth at least $6 (k + 12\eg(G)) (\eg(G)+1)$,
  %then there is a sequence of contractions that brings $G$ to
  %a partially triangulated $k \times k$
  %grid augmented with at most $\eg(G)$ additional edges.
  %
  %For any $k \geq 12 \eg(G)$,
  %If $G$ has treewidth at least $k$,
  %then there is a sequence of contractions that brings $G$ to
  %a partially triangulated $(k - 12\eg(G)) / (6 (\eg(G) + 1)) \times ditto$
  %grid augmented with at most $\eg(G)$ additional edges.
  %
  % and replace \eg(G) with g
  %
  There is a sequence of contractions that brings any graph $G$
  of genus $g$ to a partially triangulated
  $\Omega(\tw(G) / (g+1)) \times \Omega(\tw(G) / (g+1))$
  grid augmented with at most $g$ additional edges.
\end{theorem}

\begin{corollary} \label{Lovasz}
  There is a sequence of contractions that brings any graph $G$
  of genus $g$ to a partially triangulated
  $\Omega(\tw(G) / (g+1)^2) \times \Omega(\tw(G) / (g+1)^2)$ grid,
  augmented with at most $g$ additional edges
  incident only to boundary vertices of the grid.
\end{corollary}

\begin{proof}
  We take the augmented
  $\Omega(\tw(G) / (g+1)) \times \Omega(\tw(G) / (g+1))$ grid
  guaranteed by Theorem~\ref{handle},
  and find the largest square subgrid that does not contain in its interior
  any endpoints of the at most $g$ additional edges.
  This subgrid has size
  $\Omega(\tw(G) / (g+1)^2) \times \Omega(\tw(G) / (g+1)^2)$
  because there are $2 g$ vertices to avoid.
  Then we contract all vertices outside this subgrid
  into the boundary vertices of this subgrid.
\end{proof}

The main idea for proving a relation between the treewidth of a graph
and the treewidth of its dual is to relate both to the treewidth of the
radial graph, and use that the radial graph of the primal is equal to the
radial graph of the dual.

\begin{theorem} \label{graph vs. radial}
  For a 2-connected graph $G$ 2-cell embedded in a surface of genus~$g$,
  its treewidth is within an $O((g+1)^2)$ factor of the treewidth of its
  radial graph $R(G)$.
\end{theorem}

\begin{proof}
  We follow the part of the proof of Lemma \ref{map tw} establishing that
  $\tw(G) + 1 = \Omega(\tw(R \cup G)+1)$, in order to prove that
  $\tw(G) + 1 = \Omega(\tw(R)+1)$.  The differences are as follows.
  Every occurence of $R \cup G$ is replaced by~$R$.
  Instead of applying Theorem \ref{planar grid} to obtain a grid minor $K$
  and then discarding the edge deletions from the sequence to obtain a
  partially triangulated grid contraction~$K'$,
  we use Corollary \ref{Lovasz} to obtain a partially triangulated
  $\Omega(\tw(R) / (g+1)) \times \Omega(\tw(R) / (g+1))$
  grid contraction $K'$ of $R$ augmented with at most $g$ additional edges
  incident only to boundary vertices of the grid.
  Otherwise, the proof is identical, and we obtain an
  $\Omega(\tw(R) / (g+1)^2) \times \Omega(\tw(R) / (g+1)^2)$ grid contraction
  $K''$ of~$G$.
  Therefore, $\tw(G) + 1 = \Omega(\tw(R) / (g+1)^2)$.
  Because $G$ is 2-connected, $\tw(G) > 0$, so
  $\tw(G) = \Omega(\tw(R) / (g+1)^2)$.

  Now we apply what we just
  proved---$\tw(G) = \Omega(\tw(R(G)) / (g+1)^2)$---substituting $R(G)$
  for~$G$.
  (The theorem applies:
  $R(G)$ is 2-cell embeddable in the same surface as~$G$, and
  $R(G)$ is 2-connected because $G$ (and thus $G^*$) is 2-connected.)
  Thus $\tw(R(G)) = \Omega(\tw(R(R(G))) / (g+1)^2)$.
  We claim that $G$ is a minor of $R(R(G))$,
  which implies that $\tw(G) \leq \tw(R(R(G)))$
  and therefore $\tw(R(G)) = \Omega(\tw(G) / (g+1)^2)$ as desired.

  Now we prove the claim.
  Because $G$ is 2-connected,
  each face of the radial graph $R(G)$ is a diamond (4-cycle)
  $v_1, f_1, v_2, f_2$ alternating between vertices ($v_1$ and $v_2$)
  and faces ($f_1$ and $f_2$) of~$G$.
  Also,
  %because $G$ is 2-connected,
  $v_1 \neq v_2$ and $f_1 \neq f_2$.
  If we take the radial graph of the radial graph, $R(R(G))$,
  we obtain a new vertex $w$ for each such diamond,
  connected via edges to $v_1$, $f_1$, $v_2$, and~$f_2$.
  For each such vertex $w$, we delete the edges $\{w, f_1\}$
  and $\{w, f_2\}$, and we contract the edge $\{w, v_2\}$.
  The local result is just the edge $\{v_1, v_2\}$.
  Overall, we obtain $G$ as a minor of $R(R(G))$.
\end{proof}

With this connection to the radial graph in hand, we can prove the main
theorem of this section:

\begin{theorem} \label{primal vs. dual}
  The treewidth of a graph $G$ 2-cell embedded in a surface of genus $g$
  is at most $O(g^4)$ times the treewidth of the dual $G^*$.
\end{theorem}

\begin{proof}
  If $G$ is 2-connected, then by Theorem \ref{graph vs. radial},
  $\tw(G)$ is within an $O(g^2)$ factor of $\tw(R(G))$.
  Because $R(G^*) = R(G)$, we also have that
  $\tw(G^*)$ is within an $O(g^2)$ factor of $\tw(R(G))$.
  Therefore, $\tw(G)$ is within an $O(g^4)$ factor of $\tw(G^*)$.

  Now suppose $G$ has a vertex 1-cut $\{v\}$.
  Then $G$ has two strictly smaller induced subgraphs $G_1$ and~$G_2$
  that overlap only at vertex~$v$ and whose union $G_1 \cup G_2$ is~$G$.
  The treewidth of $G$ is the maximum of the treewidth of $G_1$
  and the treewidth of~$G_2$.
  (Given tree decompositions of $G_1$ and~$G_2$, pick a node in each tree
  whose bag contains~$v$, and connect these nodes together via an edge.)
  Furthermore, the dual graph $G^*$ has a cut vertex $f$ corresponding to~$v$,
  and $G^*$ similarly decomposes into induced subgraphs $G^*_1$ and $G^*_2$
  such that $G^*_1 \cup G^*_2 = G^*$ and $G^*_1$ and $G^*_2$ overlap
  only at~$f$.
  By induction, $\tw(G_i)$ is within a $c g^4$ factor of $\tw(G^*_i)$,
  for $i \in \{1,2\}$ and for a fixed constant~$c$.
  Therefore, $\tw(G) = \max\{\tw(G_1),\tw(G_2)\}$
  is within a $c g^4$ factor of
  $\max\{\tw(G^*_1),\tw(G^*_2)\}) = \Theta(\tw(G^*))$.
\end{proof}

The bound is Theorem \ref{primal vs. dual} is not necessarily the best
possible.  In particular, we can improve the bound from $O(g^4)$ to $O(g^2)$.
Instead of using Corollary \ref{Lovasz}, we can apply Theorem \ref{handle}
directly and instead modify the grid argument of Lemma \ref{map tw} to
avoid the endpoints of the $g$ additional edges.  Specifically, we stretch
the ``waffle'' of horizontal and vertical strips in the grid
connecting the $v_{i,j}$'s, so that all grid points we use for paths
avoid all rows and columns containing the endpoints of the $g$ additional
edges.  Then we can use the same argument, deleting the vertices and edges
not on the paths, and in particular deleting the $g$ additional edges,
to form the desired grid minor.

\ifabstract
\begin{theorem}
  The treewidth of a graph $G$ 2-cell embedded in a surface of genus $g$
  is at most $O(g^2)$ times the treewidth of the dual $G^*$.
\end{theorem}
\fi

\section{Conclusion}

We have proved polynomial bounds on the treewidth necessary to guarantee
the existence of an $r \times r$ grid minor for both map graphs and
power graphs, which can have arbitrarily large cliques and thus do not
exclude any fixed minor.  The techniques of our paper use approximate
max-min relations between the size of grid minors and treewidth, and
our results provide additional such relations for future use.

One of the main open problems is to close the gap between the best current
upper and lower bounds relating treewidth and grid minors.  For map graphs,
it would be interesting to determine whether our analysis is tight,
in particular,
whether we can construct an example for which the $O(r^3)$ bound is tight.
Such a construction would be very interesting because it would improve the
best previous lower bound of $\Omega(r^2 \lg r)$ for general graphs
\cite{RobSeymT94}.  We make the following stronger claim about general graphs:

\begin{conjecture}
  For some constant $c > 0$, every graph with treewidth
  at least $c r^3$ has an $r \times r$ grid minor.
  Furthermore, this bound is tight: some graphs have
  treewidth $\Omega(r^3)$ and no $r \times r$ grid minor.
\end{conjecture}

This conjecture is consistent with the belief of Robertson, Seymour, and
Thomas \cite{RobSeymT94} that the treewidth of general graphs
is polynomial in the size of the largest grid minor.
%, in the vacinity of the current best lower bound
%$\Omega(r^2 \lg r)$.

\section*{Acknowledgments}

We thank L\'aszl\'o Lov\'asz for helpful discussions about the proof
of Corollary~\ref{Lovasz}.

\small

% Decrease the space between bibliography items.
\let\realbibitem=\bibitem
\def\bibitem{\par \vspace{-1.2ex}\realbibitem}

\bibliography{Hminorfree,planar1,ourpapers}
\bibliographystyle{alpha}

\end{document}